\documentclass[12pt]{iopart}
\usepackage{setstack}

\newcommand{\Ai}{\mathrm{Ai}\,}
\newcommand{\sgn}{\mathrm{sgn}\,}
\usepackage{graphicx}

\usepackage{color}

\def\void#1{#1}

\def\myscalea{1}
\def\myscaleb{0.66}

\begin{document}

\title[Cross-phase modulation mediated pulse control with Airy pulses in optical fibers]{Cross-phase modulation mediated pulse control with Airy pulses in optical fibers}

\author{Michael Goutsoulas, Vassilis Paltoglou  and \\
 Nikolaos K. Efremidis\footnote{Author to whom any correspondence should be addressed.}}

\address{Department of Mathematics and Applied Mathematics, University of Crete, 70013 Heraklion, Crete, Greece}
\ead{nefrem@uoc.gr}

\vspace{10pt}
\begin{indented}
\item[]\today
\end{indented}

\begin{abstract}
We show that the velocity and thus the frequency of a signal pulse can be adjusted by the use of a control Airy pulse. In particular, we utilize a nonlinear Airy pulse which, via cross-phase modulation, creates an effective potential for the optical signal. Interestingly, during the interaction, the signal dispersion is suppressed. Importantly, the whole process is controllable and by using Airy pulses with different truncations leads to predetermined values of the frequency shifting. Such a functionality might be useful in wavelength division multiplexing networks.
\end{abstract}
\maketitle

\section{Introduction}

The study of curved and accelerating waves has attracted a lot of attention over the last years. This interest was triggered by the prediction and experimental observation of exponentially truncated curved Airy beams~\cite{sivil-ol2007,sivil-prl2007}. The highly desirable properties of the Airy wave (diffraction-free, self-healing~\cite{broky-oe2008,hu-ol2010}, and accelerating) have been utilized in a variety of applications from particle manipulation~\cite{baumg-np2008,zhang-ol2011}, filament generation~\cite{polyn-science2009,polyn-prl2009} and electric discharges~\cite{cleri-sa2015}, to beam autofocusing~\cite{efrem-ol2010,papaz-ol2011,zhang-ol2011}, micromachining~\cite{mathi-apl2012}, and high resolution imaging~\cite{jia-np2014,vette-nm2014}. 

In the temporal domain, accelerating pulses have been observed in the form of spatiotemporal light bullets~\cite{chong-np2010}, and utilized for supercontinuum generation in optical fibers~\cite{ament-prl2011}, for tuning the frequency of a laser pulse via the use of an Airy pulse-seeded soliton self-frequency shift~\cite{hu-prl2015}, as well as for high-aspect ratio machining of dielectrics~\cite{gotte-optica2016}. The presence of third-order dispersion can either increase the acceleration of Airy pulses~\cite{besier-pre2008} or cause inversion and tight focusing~\cite{dribe-ol2013}. Causality effects of acccleraring pulses have also been discussed~\cite{kamin-oe2011}.

\void{Of main interest, in connection to this work, is the possibility to modify the properties of an optical signal via the action of cross-phase modulation (XPM) of a nonlinear control pulse.}
The interaction regimes due to XPM of two pulses and their dynamics have been studied in~\cite{agraw-pra1989}. 
The utilization of XPM allows for an optical event horizon or the generation of an optical refractive index barrier~\cite{philb-science2008, skrya-pre2005,webb-nc2013}.
Different applications of XPM in all-optical devices have been proposed and observed~\cite{sharp-ptl2002,laroc-el1990,bogri-ptl2007,qiu-ptl2010}. 
The non-instantaneous Raman nonlinearity can cause acceleration of a soliton that can be utilized to trap an optical pulse~\cite{gorba-pra2007,nishi-ol2002}.
Our proposal relies on the use of a nonlinear Airy pulse. In this respect, analysis of the nonlinear dynamics of Airy pulses~\cite{fatta-oe2011,hu-ol2013} and their Painl\'eve II generalizations~\cite{giann-pla1989,kamin-prl2011} have been examined in the literature. Low power Airy pulses have been utilized to achieve control of a solitary wave~\cite{cai-oc2014}. In the spatial domain, a two-dimensional Airy beam lattice has been employed to achieve optical routing~\cite{rose-apl2013}. 

In this work we show that an Airy (control) pulse can be utilized to modify the properties of a Gaussian (signal) pulse. These two pulses have different wavelengths and interact nonlinearly through XPM. In particular, the Airy pulse is nonlinear and through XPM generates an effective potential for the signal. Due to their interaction, the signal pulse (which is linear or moderately nonlinear) is trapped by the effective potential and thus follows the same 
\void{(temporally translated)}
accelerating parabolic trajectory. After the interaction, the group velocity and thus the spectrum of the signal pulse are shifted. More importantly, this process is dynamic, and thus by using Airy pulses with different truncations leads to different predetermined values for the spectral shift. Such a functionality might be useful for routing pulses between different channels in wavelength division multiplexing (WDM) networks. In addition, during the interaction and due to the guiding from the control pulse, the dispersion of the signal is significantly suppressed.

\section{Modeling of the system}
We start our analysis by considering two pulses of different colors propagating in an optical fiber. Their dynamics can be modeled by the following coupled nonlinear Schrd\"oginer equations
\[
i\left(
\frac{\partial A_1}{\partial Z}+\frac 1{v_{g1}}
\frac{\partial A_1}{\partial T} 
\right)
-
\frac{\beta_{21}}2
\frac{\partial^2A_1}{\partial T^2}
+\gamma_1(|A_1|^2+2|A_2|^2)A_1=
i\frac{\beta_{31}}{6}\frac{\partial^3A_1}{\partial t^3}
+T_R\frac{\partial |A_1|^2}{\partial t}A_1
\]
\[
i\left(
\frac{\partial A_2}{\partial Z}+\frac 1{v_{g2}}
\frac{\partial A_2}{\partial T} 
\right)
-
\frac{\beta_{22}}2
\frac{\partial^2A_2}{\partial T^2}
+\gamma_2(|A_2|^2+2|A_1|^2)A_2=
i\frac{\beta_{32}}{6}\frac{\partial^3A_2}{\partial t^3}
+T_R\frac{\partial |A_2|^2}{\partial t}A_2
\]
where $A_1$, $A_2$ are the amplitudes of the electromagnetic field, $Z$ and $T$ are the spatial and temporal coordinates, $v_{gj}$ are the group velocities, $\beta_{2j}$, $\beta_{3j}$, are the second and third order dispersion coefficients, $\gamma_j = n_2\omega_j/(cA_\mathrm{eff})$ are the Kerr nonlinear coefficients ($j=1,2$) with $A_\mathrm{eff}$ being the effective mode area, $n_2$ is the nonlinear index parameter, $\omega_j$ is the respective optical frequency, and $c$ the speed of light. Finally, $T_R$ is the coefficient of the Raman nonlinearity. Since the absolute value of the difference between the two frequencies is relatively small $|\omega_1-\omega_2|\ll\omega_1,\omega_2$ we can approximate $\gamma_1\approx\gamma_2\approx\gamma$. The above system of equations can be written in a normalized dimensionless form as
\begin{equation}
i \frac{\partial u}{\partial z}-\frac{\beta_u}{2}
\frac{\partial^2u}{\partial t^2}+(|u|^2+2|v|^2)u =
i\Delta_u\frac{\partial^3u}{\partial t^3}+
\tau_R\frac{\partial |u|^2}{\partial t}u,
\label{eq:nls1} 
\end{equation}
\begin{equation}
i\left(\frac{\partial v}{\partial z}+\alpha\frac{\partial v}{\partial t}\right)
-\frac{\beta_v}{2}
\frac{\partial^2 v}{\partial t^2}+(|v|^2+2|u|^2)v  =
i\Delta_v\frac{\partial^3v}{\partial t^3}+
\tau_R\frac{\partial |v|^2}{\partial t}v.
\label{eq:nls2}
\end{equation}
where $\{u,v\}=\sqrt{P_0}\{A_1,A_2\}$ are the field amplitudes, $P_0$ is a scaling in the power, $z=Z/Z_0$ the spatial and  $t=(T-Z/v_{g1})/T_0$ the temporal (retarded) coordinate with $\beta_u=Z_0\beta_{21}/T_0^2$, and  $\alpha=Z_0(v_{g1}-v_{g2})/(T_0v_{g1}v_{g2})$. By defining the dispersion lengths $L_{D,1} = T_0^2/|\beta_{21}|$, $L_{D,2}=T_0^2/|\beta_{22}|$, and the nonlinear length $L_{NL}=Z_0=1/(\gamma P_0)$ we obtain 
\[
\beta_u=\frac{\sgn(\beta_{21})L_{NL}}{L_{D,1}},\quad 
\beta_v=\frac{\sgn(\beta_{22})L_{NL}}{L_{D,2}},
\] 
$\alpha=z_0(v_{g1}-v_{g2})/(T_0v_{g1}v_{g2})$, $\tau_R=T_R/T_0$, $\delta_u=\beta_{31}/(6|\beta_{21}|T_0)$, $\delta_v=\beta_{32}/(6|\beta_{22}|T_0)$, $\Delta_u=|\beta_u|\delta_u$, and $\Delta_v = |\beta_v|\delta_v$. 
Here, we restrict ourselves to group velocity matching conditions and set $\alpha=0$.

Since minor differences in the profile are not important we choose a Gaussian signal pulse
\[
v=F_s\exp[-(t-t_0)^2/\gamma_s^2], 
\]
where $\gamma_s$ is the pulse width, $F_s$ is its maximum amplitude, and $t_0$ is a temporal translation of the pulse with respect to the control pulse. In addition, the control pulse is
\[
u=F_c' \Ai(\gamma_c t)\exp(\alpha_c t), 
\]
where $\alpha_c$ is an exponential apodization to make its energy finite (and also determines the total extend of the Airy pulse), $\gamma_c$ characterizes the width of the main lobe of the pulse, and $F_c'$ is its amplitude. Note that since the maximum value of the Airy function is $\sigma\approx0.535656$, we define $F_c'=F_c/\sigma$, where $F_c$ is the actual maximum amplitude of the control pulse. The Airy pulse, even in the moderately strong nonlinear regime, follows the parabolic trajectory $t=\beta_u^2\gamma_c^3z^2/4$ (see~\cite{sivil-ol2007}) and decelerates at is propagates with $v_g=2/(\beta_u^2\gamma_c^3z)$. On the other hand, Airy pulses of the form $\Ai(-\gamma_ct)$ accelerate during propagation. More importantly there is direct mapping between these two regimes and thus is redundant to study both of these cases separately. For this reason in the rest of the paper we restrict ourselves to the case of decelerating Airy pulses. The generation of an exponentially truncated Airy pulse is possible due to the fact that its Fourier transform is a Gaussian with a cubic phase~\cite{sivil-ol2007}. The experimental procedure relies on the decomposition of the pulse spectrum by using a grating. Then by applying lenses and a cubic phase mask the Airy pulse can be generated in the Fourier space~\cite{chong-np2010,weine-rsi2000}.

We can identify four different regimes in the dynamics associated with the combinations in the signs of the dispersion coefficients ($\beta_u$, $\beta_v$). An additional parameter is the peak power of the two pulses. Obviously, the control pulse needs to be nonlinear in order to have a measurable effect in the dynamics of the signal pulse via the action of XPM. Furthermore, we select the signal pulse to be linear or weakly nonlinear. If the signal is linear its dynamics do not affect the behavior of the control pulse. 
However, as we will see, even if the signal pulse is in the moderately nonlinear regime, its dynamics do not significantly effect the Airy pulse due to its self-healing properties. 

\begin{figure}
\begin{center}
\includegraphics[width=\myscalea\columnwidth]{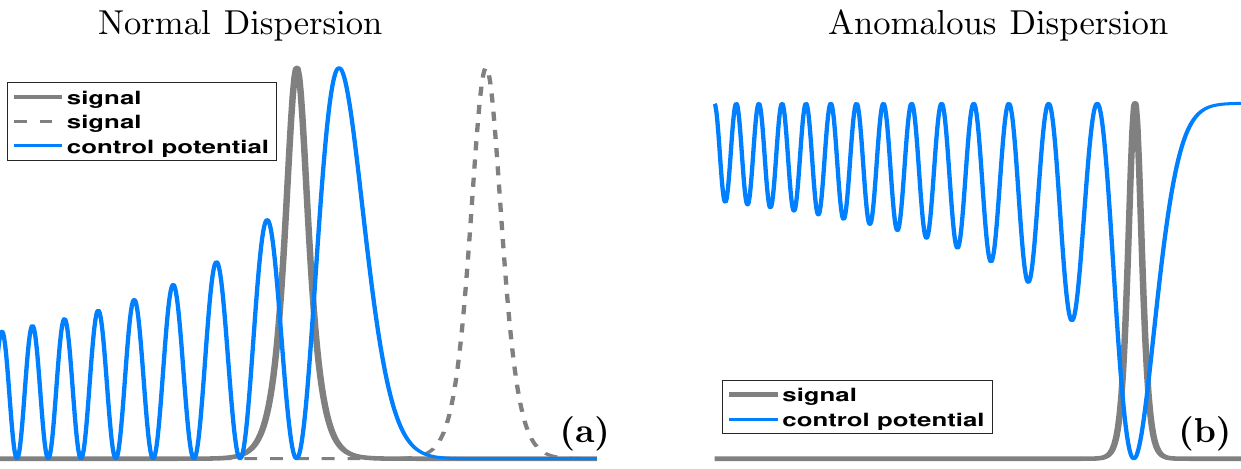} 
\caption{The Airy pulse creates an effective potential (solid blue curve) for the signal pulse (gray solid and dashed-dotted curves). (a) If the signal propagates in the region of normal dispersion then the effective potential is proportional to the power of the Airy pulse. Thus the signal can be either ``dragged'' (solid gray curve) or ``pushed'' (dashed dotted gray curve) by the potential. (b) If the signal propagates in the region of anomalous dispersion then the effective potential is inversely proportional to the power of the Airy pulse and the signal can be ``dragged'' when it is located at potential mimima.}
\label{fig:pos}
\end{center}
\end{figure}
Since the Airy pulse needs to have relatively strong nonlinearity, the selection of the sign of $\beta_u$ can significantly affect its dynamics. As it is known~\cite{fatta-oe2011} in the region of anomalous dispersion \void{accelerating} Airy pulses can suffer from soliton shedding. 
The main problem in this case is that the resulting maximum power of the soliton can be larger that the maximum power of the main Airy lobe and thus (the soliton) significantly interferes in the dynamics. Specifically, if the signal lies in the region of anomalous dispersion it can get trapped by the potential created by the soliton rather than by the Airy pulse. On the other hand, if the signal lies in the region of normal dispersion the potential barrier created by the soliton can destabilize the dynamics of the signal pulse. Thus, in the present work we restrict ourselves to the optimal case of Airy pulses propagating in the region of normal dispersion.

The sign of the dispersion $\beta_v$ is also significant because it determines the kind of potential that the signal it is subjected to. 
Let us denote as the first minimum (or maximum) of the Airy function the rightmost one and then continue counting along the negative axis.
In the same fashion we can enumerate the local maxima of the Airy function (the first one being the global maximum).
As can be seen in Fig.~\ref{fig:pos} the Airy pulse, due to XPM, creates an effective potential for the signal. In the region of normal dispersion, this potential is proportional to the power of the Airy pulse and, thus, the signal can be trapped if it is located in an Airy minimum. Furthermore, as the control pulse is accelerating, we expect the signal to follow the same kind of trajectory. We call this behavior ``dragging''. As we shift to higher numbers of minima the potential barrier contrast is decreased. In addition the frequency of the oscillations increases making the barrier ``thinner''. In this respect the first minima are preferable. 
In addition, the signal pulse can be located to the right of the Airy pulse. Thus, as the Airy pulse accelerates, it can ``push'' the signal pulse [Fig.~\ref{fig:pos}(a) dashed line]. On the other hand, if the signal propagates in the region of anomalous dispersion, the effective potential created by the Airy pulse is inversely proportional to its power. Thus, in order to facilitate the dragging behavior, the signal pulse should be located to a maximum of the Airy function. The first lobe creates the strongest potential well, and thus is preferable. However, even subsequent lobes of the Airy pulse can be utilized to drag the optical signal.

\begin{figure}[t]
\begin{center}
\includegraphics[width=\myscalea\columnwidth]{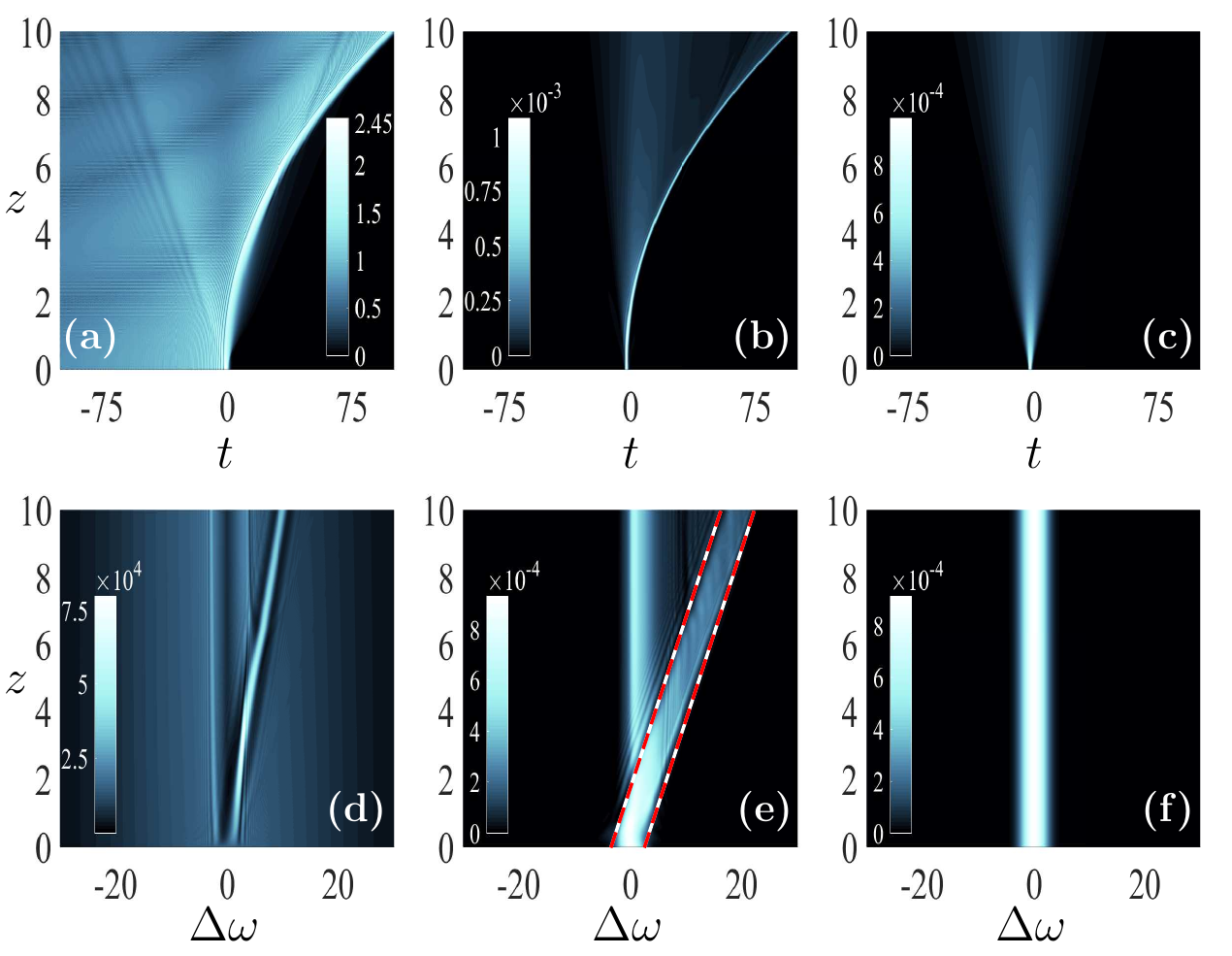}
\caption{Dynamics of the pulse amplitude (first row) and the corresponding power spectrum (second row) for $\beta_u=2$, $\beta_v=1$, $\tau_R=3/2000$, and $\Delta_u=\Delta_v=1/1200$. The first two columns depict the control Airy ($|u|$) and the Gaussian signal ($|v|$) pulses, whereas in the third column the dynamics of the signal pulse $|v|$ for $u=0$ is shown. The signal pulse is effectively linear with $F_s=0.001$ $\gamma_s=0.651$ and $t_0=-2.3482$ (positioned between main and secondary lobe), while for the control pulse $F_c=2.5$, $\gamma_c=1$, and $ \alpha_c=0.001$. 
} 
\label{fig:NN}
\end{center}
\end{figure}

\section{Numerical results}
All our simulations were carried out by utilizing a 4th order split-step Fourier scheme~\cite{agrawal-nfo}. In addition, 
we select typical values for $\beta_{3j}=0.1$ ps$^3$/km, $T_R=3$ fs, $|\beta_{21}|=20$ ps$^2$/km, and $T_0=2$ ps. The dispersion of the signal pulse is then obtained though $\beta_{22}=\beta_v\beta_{21}/\beta_u$.
Note that in all the cases studied below, the Airy pulse propagates in the region of normal dispersion, and the effect of higher order terms (Raman nonlinearity and third order dispersion) is insignificant. By comparing the simulation results produced with and without these terms we see that they are nearly identical. This can be explained by the fact that in the region of normal dispersion the main lobe of the Airy pulse has the tendency to effectively ``defocus''. In addition, following a ray optics approach, the rays that contribute to the generation of the Airy pulse originate from the low intensity tails of the Airy pulse and are different at each propagation distance. 
On the other hand, in the case where the Airy pulse propagates in the anomalous dispersion regime (which in not examined in this work) we can see the soliton (generated by shedding from the Airy pulse) decelerating due to the Raman effect.

First we examine the case where both pulses propagate in the normal dispersion regime. In the results shown in Fig.~\ref{fig:NN} the signal pulse is linear. Thus, if the control pulse is absent, the signal maintains its spectrum and its temporal profile disperses as it propagates [Fig.~\ref{fig:NN}, 3rd column]. In the first two columns of Fig.~\ref{fig:NN} we see the dynamics when both pulses co-propagate inside the fiber. We note that the spectrum of the Airy pulse changes during propagation due to its nonlinear self-action~\cite{hu-ol2013}. This mainly involves frequencies in the central part of the spectrum that are blue-shifted. However, the spectral tails, representing the driving force of the Airy pulse are not affected up to $z=10$. In~\ref{fig:NN}(b) we can see the dynamics of the signal pulse, which is originally positioned between the main and the secondary lobe. Due to the presence of the effective potential from the control the optical signal is dragged along the parabolic trajectory. We note that as the signal accelerates a part of its energy turns into low amplitude radiation. In addition, the dispersion of the signal is strongly suppressed [compare Figs.~\ref{fig:NN}(b), (c)]). In~\ref{fig:NN}(e) we depict the power spectrum $|\tilde v(\omega)|^2$ of the signal pulse. At the initial stages of propagation, we clearly see a splitting in the spectrum: The left branch, associated with the radiation, remains almost invariant during propagation. The part of the spectrum that linearly translates to the right is associated with the accelerating signal. Taking into account the group velocity of the Airy pulse
\[
v_g = \frac{2}{\beta_u^2\gamma_c^2z}
\]
which is equalized to the group velocity of the control pulse
\[
v_g = \frac1{\beta_v\Delta\omega}
\]
we obtain the frequency shift of the signal pulse as it propagates through the fiber 
\begin{equation}
\Delta\omega = \frac{\beta_u^2\gamma_c^3z}{2\beta_v}.
\label{eq:domega}
\end{equation}
We see that, as long as the signal is dragged by the control pulse, its frequency is shifting linearly with the propagation distance. The red lines in Fig.~\ref{fig:NN}(e) are translations of Eq.~(\ref{eq:domega}) to depict the spectral range of the accelerating signal. Note the excellent agreement between theoretical and numerical results.

\begin{figure}[t]
\begin{center}
\includegraphics[width=\myscaleb\columnwidth]{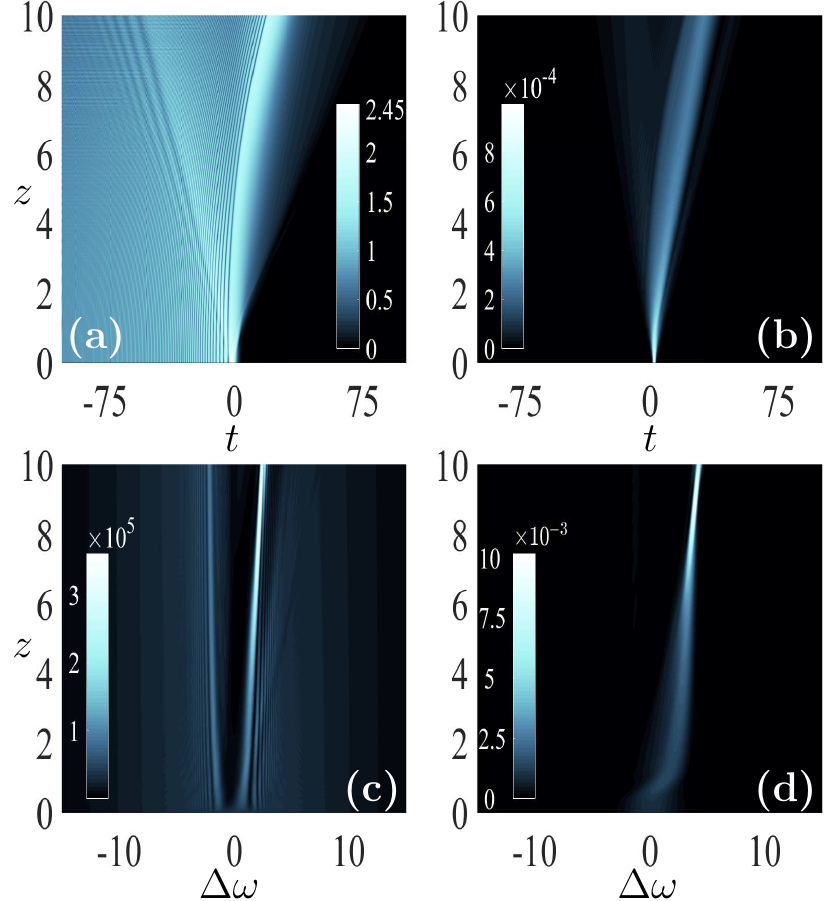} 
\caption{Dynamics of the control and the signal pulse (first row) and the corresponding spectra (second row) when the signal is launched after the Airy pulse ($t_0=2.3482$) for $\beta_u=2$, $\beta_v=1$, $\tau_R=3/2000$, and $\Delta_u=\Delta_v=1/1200$. The signal pulse propagates under essentially linear conditions with $F_s=0.001$, $\gamma_s=0.651$, and $t_0=2.3482$, while for the control pulse $F_c=2.5$, $\gamma_c=0.65$, and $\alpha_c=0.001$.  
} 
\label{fig:NNF}
\end{center}
\end{figure}
In addition to the aforementioned case, the signal pulse can be launched after the Airy pulse [Fig.~\ref{fig:pos}(a) dashed-dotted curve]. Thus, during propagation the signal is ``pushed'' from one side from the Airy pulse, but from the other side it can still disperse. In Fig.~\ref{fig:NNF} we depict the dynamics of the signal and the control pulses in the real and in the Fourier space. As we can see, the signal continues to disperse during acceleration. Perhaps the most interesting feature in the dynamics is that the spectrum of the signal gets significantly compressed as it propagates.

\begin{figure}[t]
\begin{center}
\includegraphics[width=\myscalea\columnwidth]{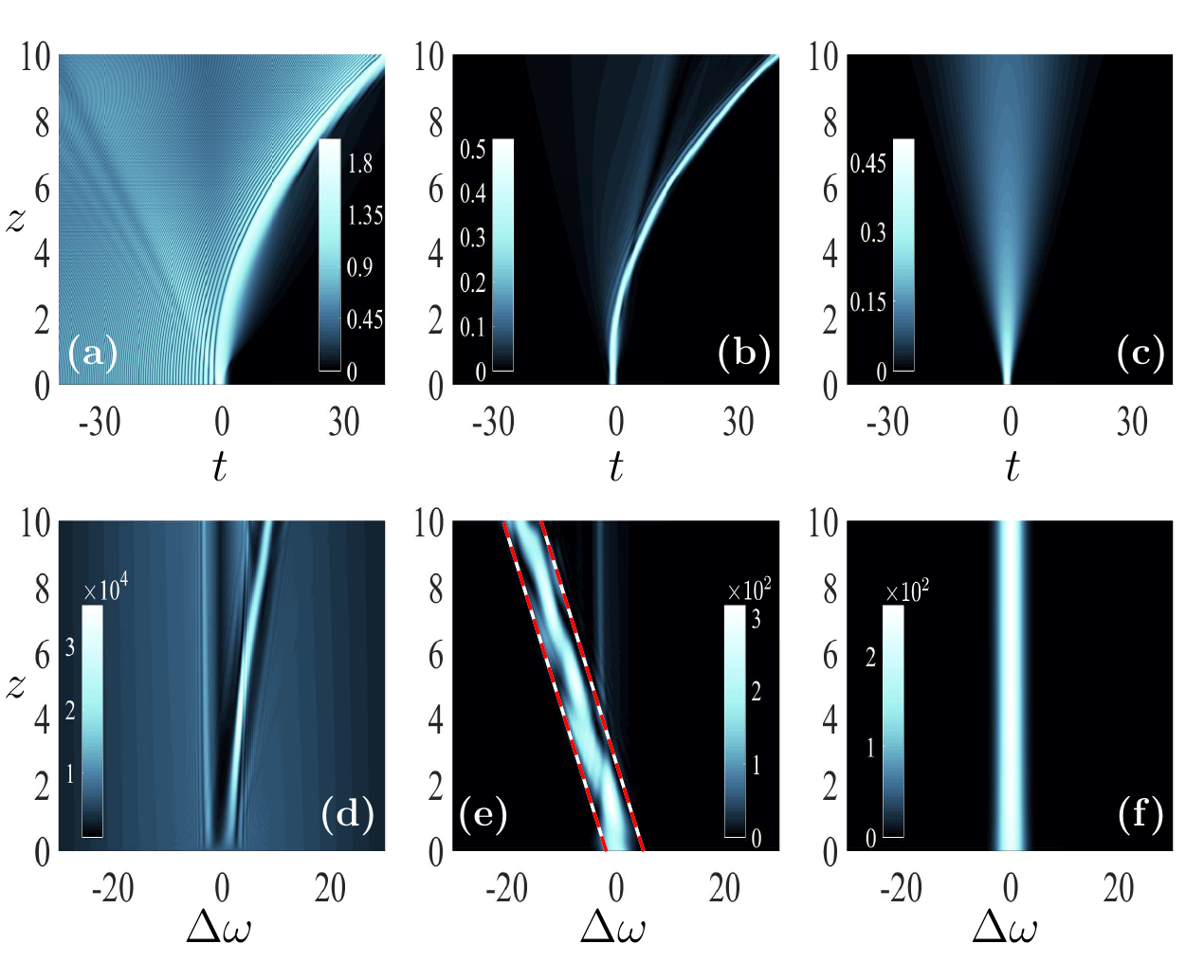} 
\end{center}
\caption{
Dynamics of the pulse amplitude (first row) and the corresponding power spectrum (second row) for $F_s=0.5$, $\gamma_s=0.651$, $t_0=-0.8789$ [the signal positioned in the main Airy lobe\void{, see Fig.~\ref{fig:pos}(b)}], $F_c=2$, $\gamma_c=1.2$, $\alpha_c=0.001$, $\beta_u=1$, $\beta_v=-1/2$, $\tau_R=3/2000$, and $\Delta_u=\Delta_v=1/2400$.}
\label{fig:NA}
\end{figure}

\begin{figure}[t]
\begin{center}
\includegraphics[width=\myscalea\columnwidth]{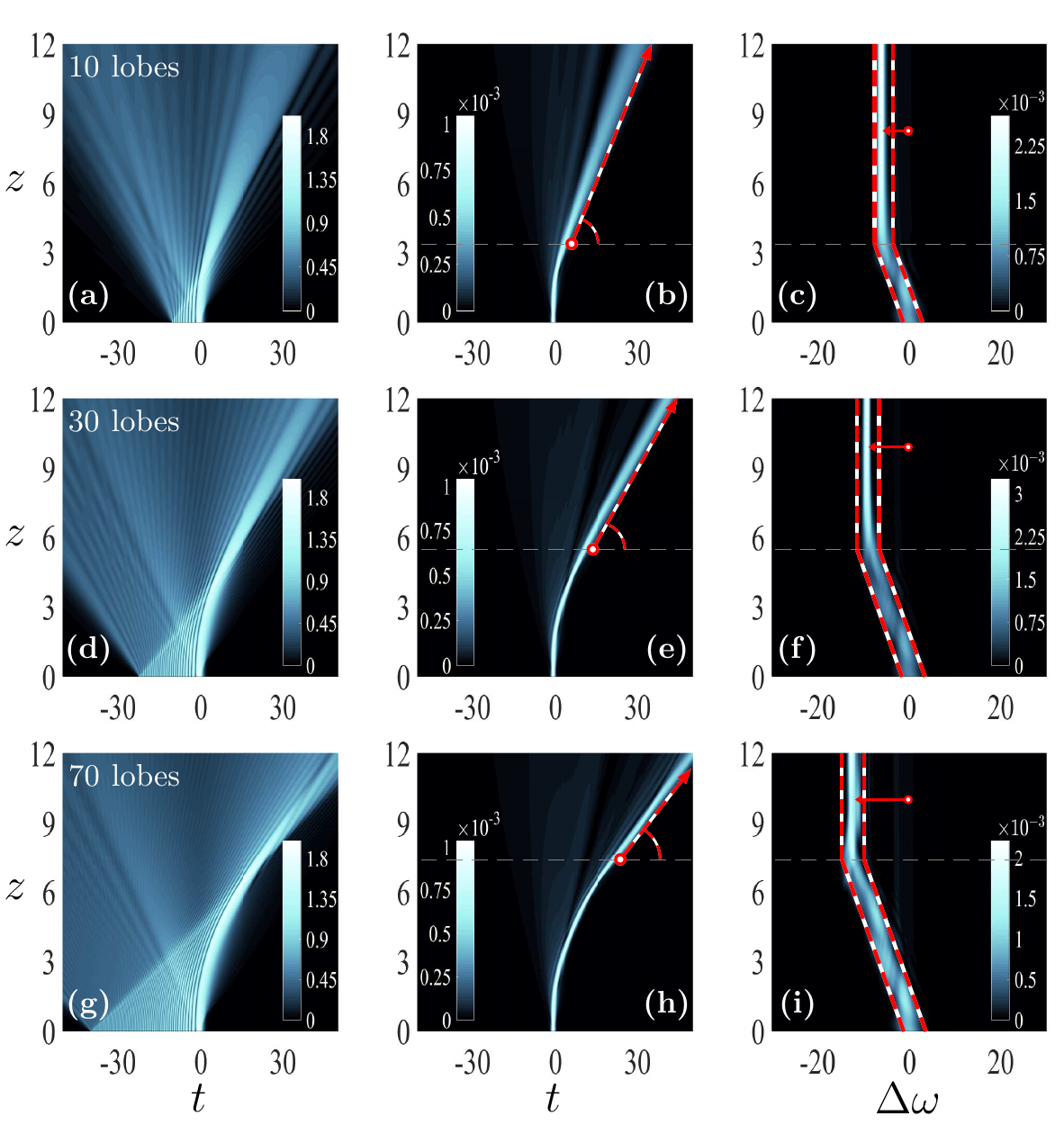}
\caption{In the three rows the interaction dynamics of an Airy pulse with $10$, $30$, and $70$ lobes with a Gaussian signal is depicted. The three columns show the amplitude of the Airy pulse $|u|$, the signal pulse $|v|$ and its power spectrum $|\tilde v|^2$. The signal is essentially linear and $\gamma_s=0.651$, $F_s=0.001$, $t_0=-0.8789$ while for the control pulse $F_c=2$, $\gamma_c=1.2$ and $\alpha_c=0.001$. The parameters are $\beta_u=1$, $\beta_v=-1/2$, $\tau_R=3/2000$, and $\Delta_u=\Delta_v=1/2400$.}
\label{fig:GD}
\end{center}
\end{figure}

Next, we examine the case where the signal pulse propagates in the anomalous dispersion regime. We select the signal maximum to be positioned at the main lobe of the Airy pulse [Fig.~\ref{fig:pos}(b)]. The power of the signal is relatively strong but not strong enough to generate a soliton, as depicted in Fig.~\ref{fig:NA}(c) where the dynamics of the signal in the absence of the control is shown. In the first two columns of Fig.~\ref{fig:NA} we see the dynamics when both pulses co-propagate inside the fiber. In this case, the signal is nonlinear and thus is expected to, in principle, affect the control via XPM. However, since the nonlinearity of the signal is not very strong we do not observe any significant signs of this interaction in the control pulse. This is demonstrated in Fig.~\ref{fig:GD} where the parameters are the same, except that the signal is linear. 

The potential to shift the frequency of a signal pulse might be used in optical networks to achieve, for example, all-optical switching between different WDM channels. In order to archive such a utilization we use Airy pulses that are truncated (their amplitude becomes zero) after a specific numbers of lobes. The larger the number of lobes of the control pulse, the longer the interaction between the two pulses is going to last, leading to a bigger value in the frequency shift of the signal. The number of lobes of the Airy pulse can be designed so that, after the interaction takes place, the signal pulses are switched to specific channels of a WDM network. An example is shown in Fig.~\ref{fig:GD}, where Airy pulses with $10$, $30$, and $70$ lobes are selected. We clearly see that after the interaction with the control pulse, and depending on the number of lobes, the signal is associated with a different group velocity and thus with a different frequency.

\section{Conclusions}

In conclusion, we have shown that the frequency and thus the velocity of a signal pulse can be controllably shifted by the use of an Airy pulse via XPM. More importantly, the whole process is controllable and by using Airy pulses with different truncations can lead to predetermined values of the signal frequency shift. Such a functionality might be useful in wavelength division multiplexing networks.

\section{Acknowledgments}

M. G. is supported from the Greek State Scholarships Foundation (IKY). N.K.E. is supported by the Erasmus Mundus NANOPHI Project (contract number 2013-5659/002-001).

\section{References} 

\newcommand{\noopsort[1]}{} \newcommand{\singleletter}[1]{#1}

\end{document}